\documentclass[10pt,journal]{IEEEtran}

\usepackage{setspace}
\usepackage{cite,graphicx,amsmath,amssymb}
\usepackage{subfigure}
\usepackage{mdwmath}
\usepackage{mdwtab}
\usepackage{balance}
\usepackage{xcolor}
\usepackage{bm}
\usepackage{amsthm}
\usepackage{threeparttable}
\usepackage{algorithm}
\usepackage{algorithmic}
\usepackage{multirow}
\usepackage{flafter}
\usepackage{makecell}
\usepackage{diagbox}
\usepackage{graphicx,epstopdf}
\usepackage{epsfig}	

\usepackage{booktabs}   
\usepackage{tabularx}   
\usepackage{multirow}   
\usepackage{ragged2e}   

\graphicspath{{./src/img/}}

\newtheorem{theorem}{Theorem}

\newtheorem{lemma}{Lemma}

\newtheorem{corollary}{Corollary}

\allowdisplaybreaks
\title{\huge Generative Site-Specific Beamforming for Next-Generation Spatial Intelligence}
\author{
	Zhaolin Wang, Zihao Zhou, Cheng-Jie Zhao, and Yuanwei Liu.
	\thanks{The authors are with the Department of Electrical and Computer Engineering, The University of Hong Kong, Hong Kong (e-mail: zhaolin.wang@hku.hk, eezihaozhou@connect.hku.hk, chengjie\_zhao@connect.hku.hk, yuanwei@hku.hk).}
	\vspace{-0.7cm}
}

\begin{document}

\maketitle

\begin{abstract}
This article proposes generative site-specific beamforming (GenSSBF) for next-generation spatial intelligence in wireless networks. Site-specific beamforming (SSBF) has emerged as a promising paradigm to mitigate the channel acquisition bottleneck in multiantenna systems by exploiting environmental priors. However, classical SSBF based on discriminative deep learning struggles: (i) to properly represent the inherent multimodality of wireless propagation and (ii) to effectively capture the structural features of beamformers. In contrast, by leveraging conditional generative models, GenSSBF addresses these issues via learning a conditional distribution over feasible beamformers. By doing so, the synthesis of diverse and high-fidelity beam candidates from coarse channel sensing measurements can be guaranteed. This article presents the fundamentals, system design, and implementation methods of GenSSBF. Case studies in both indoor and outdoor scenarios show that GenSSBF attains near-optimal beamforming gain with ultra-low channel acquisition overhead. Finally, several open research problems are highlighted.
\end{abstract}

\vspace{-0.7cm}
\section{Introduction}
Multiantenna technologies have reshaped wireless networks by exploiting spatial degrees of freedom to boost data rates and reliability without additional bandwidth \cite{10144733}. A cornerstone of multiantenna technologies is beamforming, which coordinates the antenna array to focus radiated energy toward intended users instead of spreading power indiscriminately. However, beamforming is only as good as the transmitter's channel state information (CSI), an accurate description of the instantaneous propagation conditions across the array \cite{1193803}. Obtaining timely CSI is notoriously difficult. It demands recurrent estimation, feedback, and processing that must keep pace with mobility, blockage, and the growing dimensionality of massive arrays. This overhead, together with the resulting latency and protocol burden, forms a persistent bottleneck between the theoretical promise of multiantenna systems and their practical deployment.

In 5G New Radio (NR), this bottleneck is typically managed through two beamforming frameworks \cite{Dahlman2018NR}: \emph{eigen-based beamforming (EBB)} and \emph{grid of beams (GoB)}. These frameworks represent two distinct philosophies of CSI acquisition. EBB, commonly adopted in time division duplex (TDD) systems, exploits uplink-downlink reciprocity, i.e., the base station (BS) estimates the uplink channel from pilots transmitted by the user equipment (UE) and then computes downlink beamforming weights to steer a user-specific beam. In principle, EBB can approach optimal performance because it aims to recover the complete channel across the entire array. However, in practice, such accuracy requires substantial pilot resources and heavy baseband processing, and the pilot length typically scales with the array dimension. The resulting overhead reduces the fraction of each coherence block available for payload transmission and can increase latency. By contrast, GoB acquires CSI implicitly using a finite beamforming codebook. The BS sweeps reference signals over predefined beams, and the UE feeds back the index of the best beam according to measurements such as the reference signal received power (RSRP). GoB reduces signal processing complexity and feedback granularity, but its adaptivity is limited by the finite beam set. Beam misalignment can occur when the UE lies between beams, and frequent beam sweeping is needed in high-mobility scenarios. Moreover, pilot overhead remains non-negligible when large codebooks are needed to cover an entire sector or to refine the user-specific beams.

To address the inherent trade-off between the high overhead of EBB and the limited adaptivity of GoB, \emph{site-specific beamforming (SSBF)} has emerged as a promising approach~\cite{10634048}. The core idea is to replace repeated blind estimation with informed prediction by exploiting prior knowledge of the propagation environment. By digitizing the site (e.g., building geometry and persistent scatterers) and associating UE locations with site-dependent propagation characteristics, the BS can infer an effective beam using only coarse measurements, such as the RSRP over a small set of probing beams, instead of exhaustive sweeping or full-dimensional channel estimation. This environment-aware paradigm can substantially reduce CSI acquisition latency and protocol overhead while maintaining high beam accuracy tailored to the deployment. More broadly, it reflects a form of \emph{spatial intelligence} in wireless systems, namely the capability to exploit site-specific environmental knowledge together with lightweight real-time sensing in order to make context-aware decisions.

In this article, we propose a generative SSBF (GenSSBF) framework that leverages generative models to further unlock the potential of site-specific strategies and to improve beam inference beyond classical discriminative approaches, especially when SSBF is viewed as a multimodal structured prediction problem. We first summarize the fundamentals and key challenges of discriminative SSBF. We then present the GenSSBF framework, highlighting why generative models are particularly suitable for SSBF, how they can be adapted to beamforming tasks, and what network architectures and training methodologies are appropriate. Representative case studies are provided to illustrate the effectiveness of GenSSBF. Finally, we conclude by outlining several open research problems.

\begin{figure*}[t]
	\centering
	\includegraphics[width=0.7\textwidth]{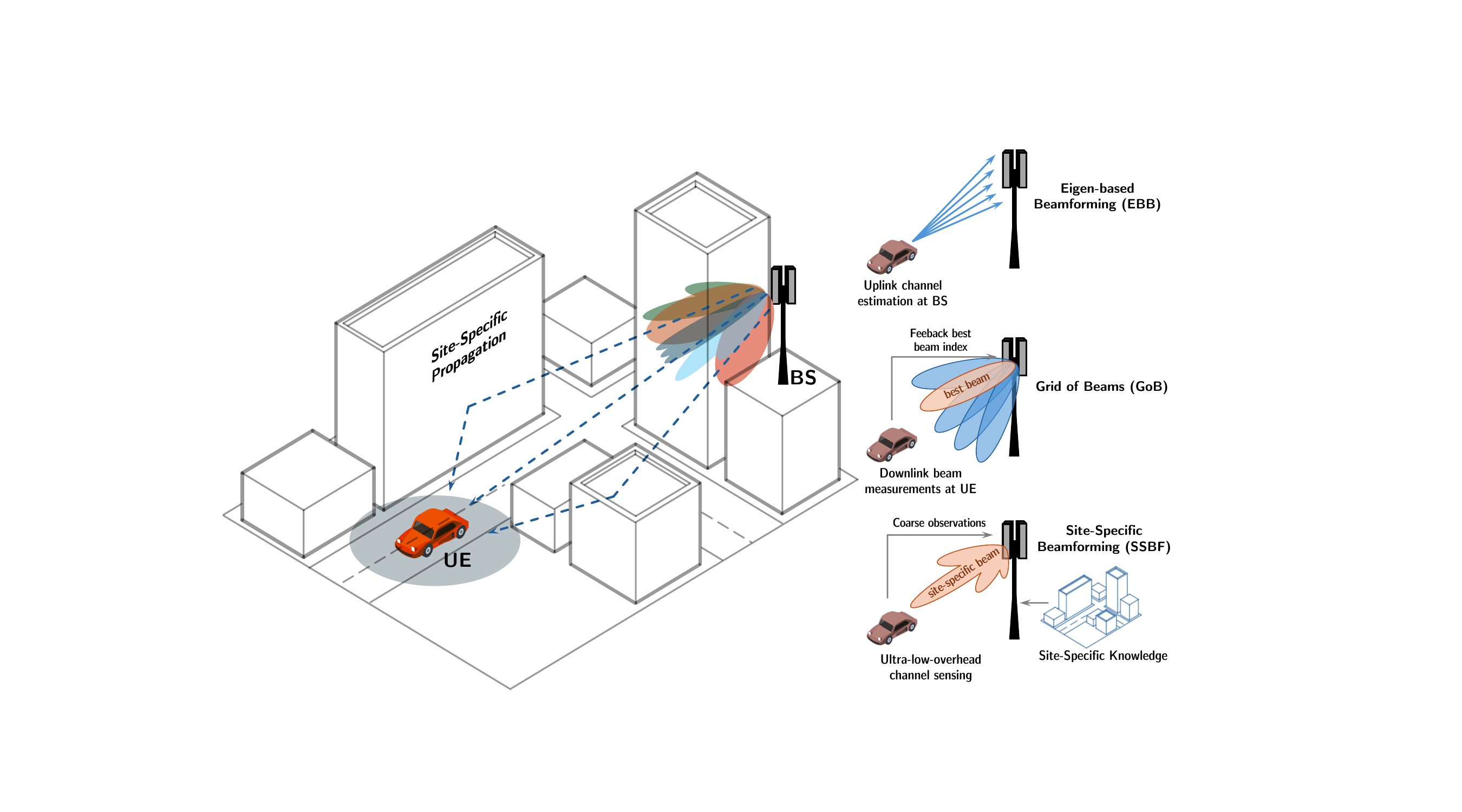}
	\caption{Illustration of the EBB, GoB, and SSBF beamforming methods. SSBF shifts the paradigm from blind channel estimation to site-specific knowledge-informed prediction.}
	\vspace{-0.2cm}
	\label{beamforming_methods}
	\vspace{-0.4cm}
\end{figure*}

\vspace{-0.45cm}
\section{Fundamentals of SSBF}
Conventional beamforming relies on repeatedly estimating high-dimensional CSI from pilots and feedback, treating the environment largely as an unknown. In contrast, SSBF exploits the fact that propagation is governed by the physical site and transceiver geometry. In particular, for a given environment, the channel or a suitable beam choice is strongly correlated with the UE location and other slowly varying contexts. SSBF therefore shifts the burden from real-time estimation to offline or semi-offline environment digitization, enabling the BS to infer an effective beam configuration using lightweight and coarse observations, as shown in Fig. \ref{beamforming_methods}. A typical SSBF system comprises three tightly coupled components that connect the physical environment to real-time beamforming, as elaborated below.

\textbf{Site-Specific Environment Database.} This component is the system \emph{memory}. It stores spatially indexed radio information about the site-specific environment. This prior knowledge enables the system to anticipate signal behavior at a given location without requiring full CSI acquisition. However, establishing such a database is non-trivial. Collecting large-scale channel measurements across an entire service area is costly and time-consuming, and often impractical in operational networks. As a result, SSBF commonly relies on \emph{digital-twin} tools to bootstrap the database by simulating site-specific propagation under realistic geometry and material assumptions. Representative examples include commercial ray-tracing platforms such as Remcom Wireless InSite \cite{Remcom_WirelessInsite} and open-source simulation frameworks such as NVIDIA Sionna \cite{sionna}, which can generate large volumes of labeled and location-indexed radio data for model training and rapid what-if evaluations. Importantly, digital twins are not a one-off asset. Real-world measurements remain essential to calibrate and continuously update the digital twin (e.g., material parameters, antenna patterns, and newly introduced scatterers), and to fine-tune the SSBF inference model so that beam predictions remain robust under temporal variations and imperfect site modeling.

\textbf{Real-Time Channel Sensing Module.} This component acts as the \emph{eyes} of the system to infer the UE's location or, more generally, its context within the site-specific environment. The inputs can come from heterogeneous sources, including position estimates, visual cues, and lightweight radio measurements. A highly effective approach is to use a small set of probing beams. Specifically, rather than sweeping a large, generic GoB codebook, the BS transmits reference signals over a compact probing set, and the UE feeds back measurements of these beams, e.g., RSRP values. Recent studies further optimize this probing codebook in a site-specific manner \cite{9690703}, i.e., training a deployment-tailored probing codebook so that a handful of measurements are maximally informative for location discriminability and beam inference under the local geometry and blockage statistics. By exploiting the strong priors embedded in the environment database, these compact probes are typically sufficient to resolve the UE context and enable accurate beam prediction, thereby reducing channel estimation complexity, pilot overhead, and signaling latency.

\begin{figure*}[t]
	\centering
	\includegraphics[width=0.72\textwidth]{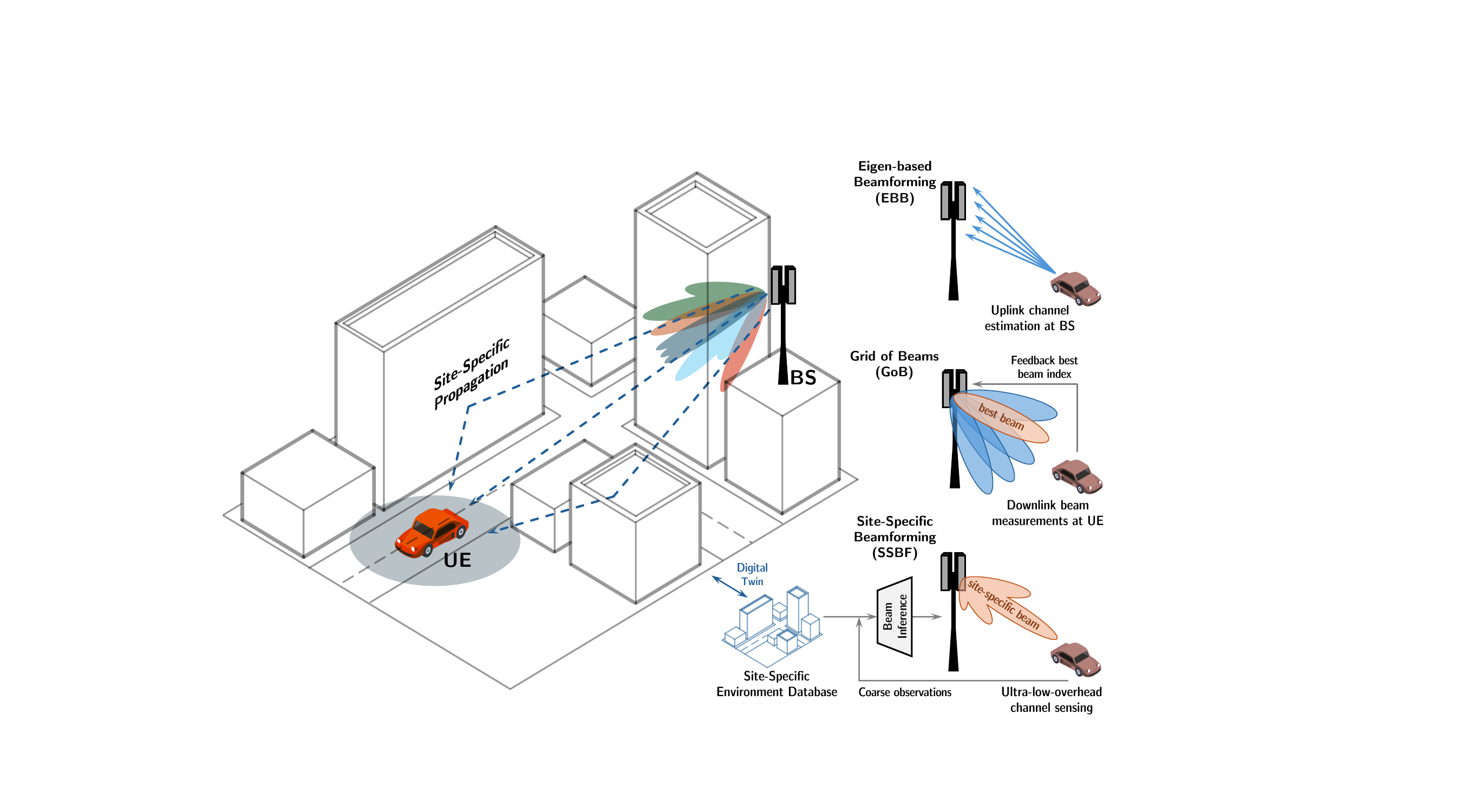}
	\vspace{-0.2cm}
	\caption{Comparison of generative and discriminative models in addressing multimodality in SSBF.}
	\label{model_comparison}
	\vspace{-0.4cm}
\end{figure*}

\textbf{Beamforming Inference Network.} This component is the system \emph{brain}. It translates the sensed context, such as UE location or coarse CSI, into a beamformer by leveraging the site-specific priors embedded in the environment database. Because the underlying mapping from coarse observations to the optimal beam depends on complex site-dependent multipath and blockage mechanisms, it is rarely amenable to an explicit analytical model. Consequently, existing SSBF implementations typically leverage discriminative deep neural networks (DNNs) to learn this mapping directly from data. In particular, current discriminative DNN-based beam inference falls into two categories \cite{10634048}. The first is \emph{codebook-based classification}, where the DNN maps coarse CSI to the index of the best beam in a pre-defined beamforming codebook. This formulation is protocol-friendly and naturally aligns with standard GoB procedures. The second is \emph{beamforming weight regression}, where the DNN learns a continuous mapping from coarse CSI to beamforming weights, aiming to synthesize a grid-free beamformer directly without being constrained to a discrete codebook. While regression can offer finer adaptivity and potentially higher performance, it typically requires careful output normalization and constraint handling, such as transmit power and hardware limitations, and is more sensitive to distribution shifts and modeling mismatch.

\vspace{-0.3cm}
\section{The Case for GenSSBF}

While discriminative models have enabled practical SSBF implementations, their limitations become evident once SSBF is cast as a \emph{multimodal structured prediction} problem. In this section, we first identify the multimodality and structured outputs of SSBF that make discriminative models struggle and discuss why the generative model is promising.

\vspace{-0.3cm}
\subsection{SSBF as Multimodal Structured Prediction}

As discussed in the previous section, the key task of SSBF is to infer an effective beamformer from a lightweight observation. Let $\bm{x}$ denote the SSBF observation, e.g., a vector of RSRP measurements over a compact probing set, and $\bm{w}$ denote the downlink beamformer, which can be either a discrete codebook index or a continuous beamforming weight vector or matrix across antennas, subcarriers, and data streams. SSBF can then be abstracted as learning a conditional mapping from $\bm{x}$ to $\bm{w}$ under a site-specific environment prior. Importantly, this is not a conventional one-to-one regression or classification problem. Instead, SSBF is inherently a \emph{multimodal structured prediction} problem, where both the multimodality and the structured output are fundamental, as elaborated in the following.

\textbf{Multimodality: one observation implies multiple plausible beams.} SSBF deliberately reduces channel sensing overhead, which makes $\bm{x}$ only partially informative about the CSI of the UE. In practice, a coarse CSI vector typically corresponds to a region of possible UE locations and propagation states, rather than a single deterministic state. This one-to-many relationship introduces multimodality in the conditional mapping, i.e., multiple beamformers can all be near-optimal under the same $\bm{x}$. Two mechanisms jointly drive this multimodality. \emph{First}, multipath propagation creates multiple viable dominant paths or clusters. Different combinations of reflections, diffractions, and blockages can yield similar coarse measurements while requiring different beamforming vectors. \emph{Second}, aggressive observation compression amplifies ambiguity. As the probing set is reduced to save pilots, feedback, and latency, distinct UE locations and propagation conditions become indistinguishable in the observation space. Consequently, SSBF is better described by a conditional distribution $p(\bm{w} \, | \, \bm{x})$ with multiple modes, rather than a single deterministic solution.

\textbf{Structured outputs: beamformers are coherent and subject to joint constraints.}
The SSBF output is structured for two reasons. \emph{First}, beamforming gain is governed by relative amplitude and phase relationships across antennas. The absolute value of an individual antenna weight is not the key quantity. Instead, the coherent structure formed by inter-antenna differences determines constructive and destructive interference in space. \emph{Second}, practical beamformers must satisfy joint feasibility constraints, such as total power constraints, per-antenna limitations, constant-modulus constraints in analog or hybrid architectures, and mutual coupling or impedance matching-related constraints in strongly coupled arrays. These constraints couple all elements of $\bm{w}$, which means that treating the outputs as independent variables can significantly increase the difficulty to synthesize the coherent and feasible beamformers.

\vspace{-0.3cm}
\subsection{From Discriminative Models to Generative Models}

Generative models, specifically conditional generative models, are a powerful approach to multimodal structured prediction problems~\cite{torralba2024foundations}. In the following, we explain why such models are particularly well-suited for SSBF, especially compared with the discriminative models, as illustrated in Fig. \ref{model_comparison}.

\textbf{Why discriminative models struggle.}
Existing SSBF solutions predominantly rely on discriminative models, either via codebook-based classification or weight regression~\cite{ 10634048}. Under multimodality, a regression model trained with point-wise losses tends to produce a compromise among multiple plausible beams. The resulting prediction can be misaligned with every true propagation mode, leading to reduced beamforming gain and fragile performance. Codebook-based classification avoids continuous averaging, but introduces a different limitation, i.e., the action space is quantized, and the beamformer must be selected from a finite set. As SSBF extends beyond single-beam selection to richer outputs such as multi-carrier, multi-stream, or multi-user beams, the label space grows rapidly, and quantization errors become increasingly pronounced. Moreover, both regression and classification are typically trained to output a \emph{single} prediction. When $\bm{x}$ is ambiguous, this forces the model to commit to one mode during inference, which yields poor beams and error propagation. Finally, encoding global coherence and physical feasibility into a purely discriminative output is non-trivial, as these unstructured methods assume independence among output elements. Therefore, prohibitively large datasets are required to accurately capture the joint dependencies across antennas, subcarriers, and users, particularly in regression approaches.


\textbf{Why generative models are needed.}
These limitations motivate the use of generative models that explicitly represent SSBF as learning a multimodal conditional distribution $p(\bm{w} \, | \, \bm{x})$, rather than a single deterministic mapping. This shift is crucial because it allows the model to retain multiple high-fidelity beam hypotheses consistent with the same coarse observation, avoiding both (i) low-quality beams produced by regression under multimodality and (ii) the need for excessively large codebooks in classification. Another key practical advantage is a \emph{generate-and-select} inference principle. Instead of producing one beam, a generative model can produce a small candidate set that spans different modes of the site-specific channel, after which the BS selects the best candidate. In this way, generative models convert unavoidable ambiguity into a controlled exploration step. Furthermore, generative models also provide a natural route to handle the structured nature of beamforming. Because a generator produces the beamformer jointly, it can better preserve coherence across antennas, subcarriers, and users, and can more naturally incorporate feasibility constraints through structured parameterizations or constrained decoding. As a result, the generative model exhibits robust generalization under limited training data and distribution shifts between digital twins and real deployments.

\vspace{-0.3cm}
\section{The Proposed GenSSBF for Next-Generation Spatial Intelligence}

As discussed in the previous section, generative models are not merely a cosmetic replacement for discriminative models in SSBF. Rather, they provide a principled solution to the intrinsic multimodality and structured-output nature of site-specific beamforming. In this section, we present the proposed GenSSBF framework and discuss its design principles, workflow, network models, and learning methods.

\begin{table*}[t]
    \centering
    \caption{Comparison of Network Models and Learning Methods for GenSSBF}
    \label{tab:comparison}
    \small 
    \setlength{\tabcolsep}{4pt} 
    \renewcommand{\arraystretch}{1.1} 

    \begin{tabularx}{\textwidth}{@{} l l X X @{}}
    \toprule
    \textbf{Category} & \textbf{Model/Method} & \textbf{Advantages} & \textbf{Disadvantages} \\ 
    \midrule
    \multirow{3}{*}{\makecell{\textbf{Network} \\ \textbf{Models}}}
        & Diffusion 
        & Multimodal distributions; Diverse beam hypotheses
        & Slow inference; High training complexity \\ 
        
        & Flow Matching 
        & Fast inference/training; Good quality-latency tradeoff
        & Complex “flow” design \\ 
        
        & LLM 
        & Fuses heterogeneous prompts; High generalization
        & Indirect generation; Resource-intensive \\ 
    \midrule
    \multirow{3}{*}{\makecell{\textbf{Learning} \\ \textbf{Methods}}} 
        & Supervised 
        & High fidelity; Simple implementation
        & High labeling cost; Simulation-to-reality gap \\ 
        
        & Self-Supervised 
        & Data-efficient; Learns inherent structure
        & Indirect output; Requires fine-tuning for alignment \\ 
        
        & Reinforcement 
        & No labels needed; Optimizes end-to-end QoS
        & Unstable convergence; Requires massive interactions \\ 
    \bottomrule
    \end{tabularx}
	\vspace{-0.5cm}
\end{table*}

\begin{figure}[t]
	\centering
	\includegraphics[width=0.4\textwidth]{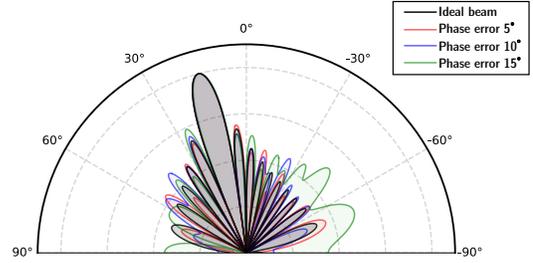}
	\caption{Beampatterns of beamformers under different phase error levels. Even small phase errors can cause significant beampattern distortion.}
	\label{phase_error}
	\vspace{-0.1cm}
\end{figure}

\vspace{-0.3cm}
\subsection{Design Principles and Workflow}

Implementing GenSSBF may appear straightforward, as it seems to require only replacing the discriminative model used for beamforming inference with a generative model. However, such a replacement is not trivial. Unlike conventional generation tasks, GenSSBF necessitates unique design principles, which are outlined as follows:
\begin{itemize} 
	\item \textbf{High-Fidelity Generation:} Unlike conventional generative tasks such as image synthesis, where minor pixel perturbations are often imperceptible, beamforming is inherently phase-sensitive, as illustrated in Fig. \ref{phase_error}. The phase of each antenna element explicitly governs whether signals combine constructively or destructively at the receiver. Consequently, GenSSBF models must prioritize phase sensitivity, ensuring that the generated beamformers accurately preserve the coherent phase relationships required to maximize array gain and suppress interference.
	\item \textbf{Robust Conditioning Mechanism:} The condition fed into the generative model acts as a \emph{wireless prompt} that guides the generation process. To ensure accurate generation in complex wireless environments, the design of the conditioning mechanism should consider: 1) how to effectively capture site-specific information in the condition embedding; 2) how to integrate multimodal data sources; 3) how to inject the condition into the model; and 4) how to design a robust conditioning framework against noise and missing data. There is an inherent trade-off here, i.e., increasing the richness of the condition improves accuracy but also increases sensing overhead.
	\item \textbf{Practicality and Compatibility:} This consideration depends on both the model architecture and the workflow design. In particular, the GenSSBF system should be fully compatible with existing 5G and 6G frameworks, thereby minimizing the need for radical standardization changes or new signaling formats. Furthermore, the ultimate objective is to generate beams that are feasible for practical hardware. Consequently, practical constraints must be incorporated into the model architecture or the inference process.
\end{itemize}

\begin{figure}[t]
	\centering
	\includegraphics[width=0.5\textwidth]{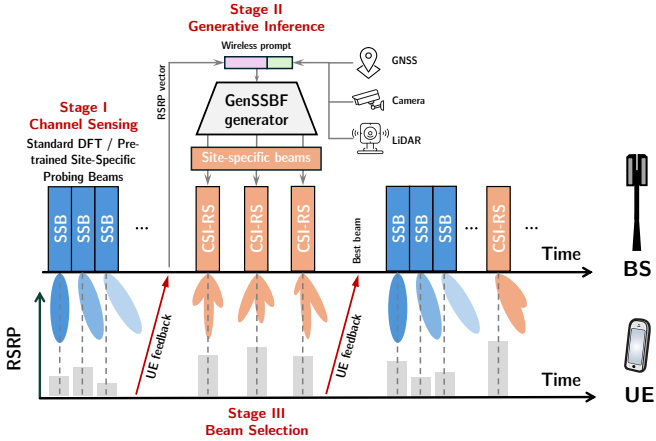}
	\caption{Reference workflow of GenSSBF in conjunction with downlink reference signals in 5G NR protocols \cite{Dahlman2018NR}.}
	\label{workflow}
	\vspace{-0.1cm}
\end{figure}

A key advantage of GenSSBF is its ability to bridge the gap between the low overhead of GoB and the high performance of customized beamforming via a \emph{generate-and-select} workflow. This approach converts the model's probabilistic output into a protocol-compliant procedure compatible with standards. The workflow proceeds in three stages, as illustrated in Fig. \ref{workflow}. 
\begin{itemize}
	\item \textbf{Stage I -- Channel Sensing:} The BS transmits a standard, coarse set of probing beams using periodic signals, such as synchronization signal blocks (SSB). The UE measures the RSRP of these beams and reports them back to the BS. This low-dimensional report serves as the condition, i.e., the wireless prompt.
	\item \textbf{Stage II -- Generative Inference:} Upon receiving the wireless prompt, the BS inputs it into the GenSSBF model. Instead of predicting a single average beam, the model generates a small set of candidate beams. These candidates represent the multiple distinct modes of the channel distribution, effectively covering the most likely propagation paths compatible with the observed RSRP.
	\item \textbf{Stage III -- Beam Selection:} To identify the best candidate beam among the generated beams, the BS configures user-specific signals, such as aperiodic channel state information reference signals (CSI-RSs), using the generated beams. The UE then measures these beams and feeds back the index of the best one.
\end{itemize}

\begin{figure*}[t]
	\centering
	\includegraphics[width=0.85\textwidth]{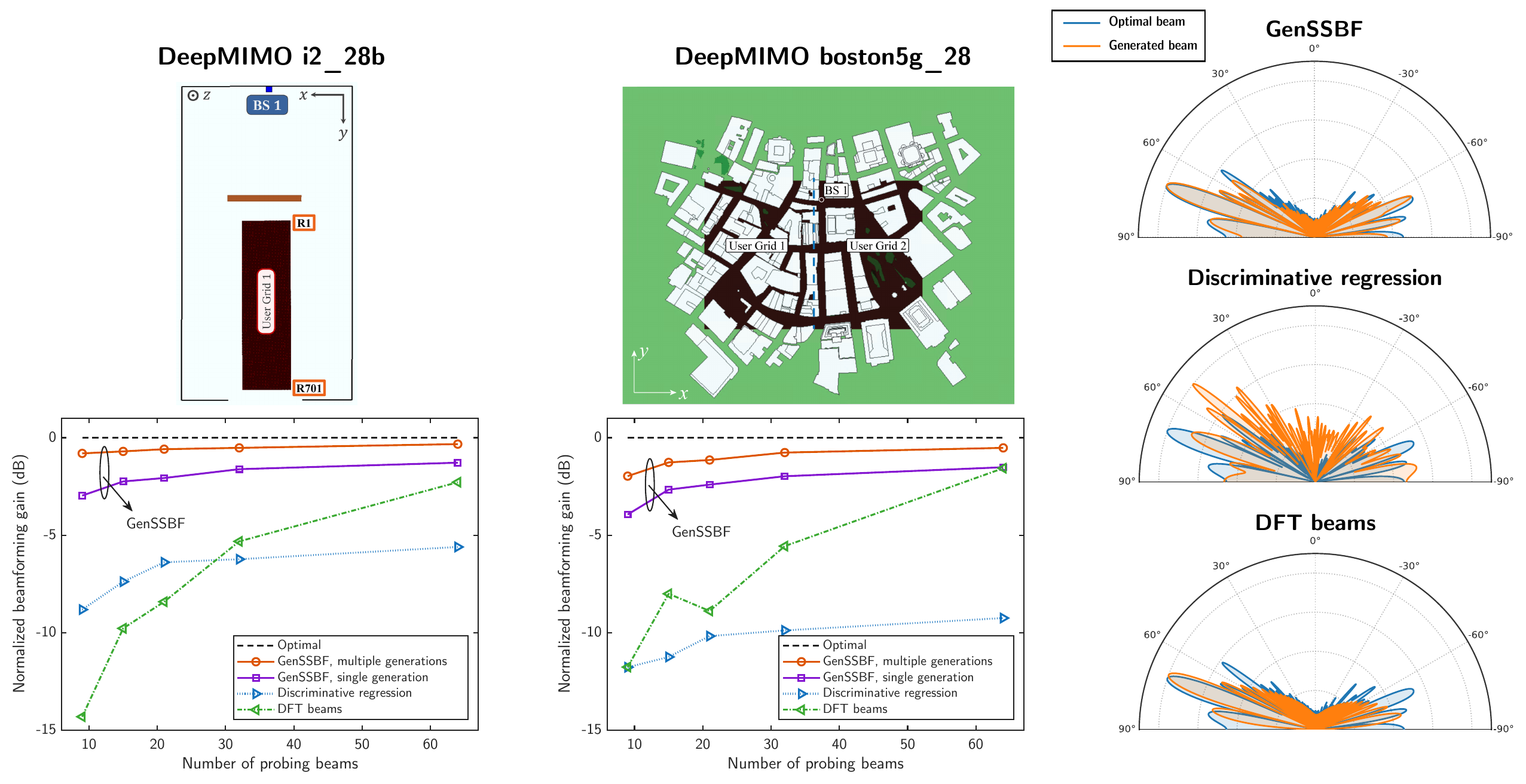}
	\vspace{-0.2cm}
	\caption{Case study results of GenSSBF in the DeepMIMO i2\_28b and boston5g\_28 scenarios \cite{alkhateeb2019deepmimo} using a conditional diffusion model. The left and middle columns show the deployment environments and the corresponding normalized beamforming gain versus the number of probing beams. The right column shows representative beampattern comparisons for GenSSBF, discriminative regression, and DFT beams against the optimal beam.}
	\label{simulation}
	\vspace{-0.5cm}
\end{figure*}

In the channel sensing stage, GenSSBF can further improve robustness by fusing heterogeneous data sources, such as RSRP measurements, GNSS coordinates, camera video streams, and LiDAR samples \cite{10719654}, into a unified \emph{multimodal wireless prompt} by leveraging its ability to process multimodal inputs. While additional modalities usually reduce uncertainty, in real-world scenarios, the data is often noisy or partially missing, meaning a multimodal wireless prompt may still be compatible with multiple valid beams. A discriminative model might average these beams, leading to poor performance. In contrast, the conditional generator absorbs these heterogeneous data sources through a shared interface, mapping them into a common latent context. This preserves the resulting ambiguity as a multimodal distribution rather than collapsing it. Consequently, when the wireless prompt is informative, the distribution concentrates on a consistent solution. When the prompt is noisy or partially missing, the model maintains diversity, relying on the generate-and-select process to resolve the ambiguity.

\vspace{-0.3cm}
\subsection{Network Models and Learning Methods}

\textbf{Network Models.} It is important to note that GenSSBF is a general framework rather than a single prescribed generator architecture. In principle, any conditional generative model that can represent the distribution of feasible beamformers given site-specific observations can be adopted. In the following, we discuss three representative candidates, each with unique advantages and trade-offs, that are particularly promising for GenSSBF: diffusion models, flow-matching models, and large language models (LLMs).
\begin{itemize}
	\item \textbf{Diffusion Models.} Diffusion models have emerged as a powerful alternative capable of capturing highly complex multimodal distributions \cite{ho2020denoising}. These models treat sample generation as an iterative denoising process that produces samples faithfully reflecting the underlying distribution. During inference, the model starts with random noise and progressively refines it under the guidance of the site-specific wireless prompt, to recover a clean beamformer. This ability to model the full conditional distribution makes diffusion models attractive for SSBF, where multiple valid beams may exist for a single wireless prompt. However, the iterative nature of diffusion sampling introduces significant latency, which poses a challenge for real-time communications where beam alignment must happen within milliseconds.
	\item \textbf{Flow-Matching Models.} Flow-matching models are a strong recent class of methods in generative deep learning \cite{lipman2023flow}. Instead of iterative denoising, they learn a continuous velocity field that deterministically transports a simple prior distribution directly to the target beamformer distribution. This results in a straighter and more efficient generation trajectory, allowing the model to synthesize optimized beamformers with significantly fewer steps than diffusion models. This combination of rapid inference and strong distributional expressiveness positions flow-matching as a leading candidate for practical low-latency GenSSBF implementations.
	\item \textbf{LLMs.} While not designed to generate beamformers directly, LLMs play a crucial complementary role in GenSSBF systems. Their strength lies in multimodal reasoning and semantic interpretation \cite{bariah2024large}. In particular, an LLM excels at processing heterogeneous data sources to refine the wireless prompts fed into generators. When combined with other models such as diffusion and flow-matching, LLMs help bridge the gap between semantic-level observations and physical-level channel realizations, enabling more
	accurate and context-aware beam generation. Furthermore, their ability to generalize across diverse scenarios also makes LLMs particularly valuable in dynamic or previously unseen environments.
\end{itemize}

\textbf{Learning Methods.} Additionally, different deployments constrain data availability and feedback pathways. GenSSBF therefore benefits from combining multiple learning methods as discussed below:
\begin{itemize}
	\item \textbf{Supervised Learning.} The most straightforward approach is supervised learning, where the model is trained to mimic a dataset of optimal beamformers labeled via exhaustive search or ray-tracing. This method yields high fidelity when high-quality data is available. However, it suffers from the simulation-to-reality gap. Specifically, models trained on synthetic digital twins may fail when deployed in real-world sites with unmodeled scatterers or hardware imperfections. Furthermore, collecting large-scale ground-truth beamforming labels in the field is often prohibitively expensive. Despite these limitations, it remains a strong baseline and essential foundation for training GenSSBF models.
	\item \textbf{Self-Supervised Learning.} To mitigate the reliance on labeled data, self-supervised learning exploits the inherent structure of wireless signals. By training the model on pretext tasks, such as predicting masked CSI entries, the network learns robust representations of the site-specific propagation environment. While this paradigm does not directly output beamformers, it serves as an excellent pre-training strategy, initializing the generator with a deep understanding of local geometry and multipath patterns before fine-tuning.
	\item \textbf{Reinforcement Learning.} Reinforcement learning offers a pathway to optimize beamformers without explicit ground-truth labels. In this framework, the generative model acts as an agent, treating the generated beam as an action and receiving a reward based on system-level metrics like spectral efficiency. This allows the model to explore the solution space and discover strategies that maximize actual network performance rather than just mimicking a dataset. When combined with conditional generative models, reinforcement learning allows the model to produce beamformers that are not only aligned with the wireless prompt but also directly optimized for a desired quality-of-service (QoS) metric.
\end{itemize}

\vspace{-0.3cm}
\subsection{Case Studies}

To concretely illustrate the GenSSBF framework, we adopt a conditional diffusion model that generates candidate beamformers from the coarse RSRP prompt, as detailed in \cite{zhou2026beam}. The model is trained in a supervised manner using paired sensing-beamforming samples constructed from DeepMIMO scenarios \cite{alkhateeb2019deepmimo}.  Specifically, we consider two DeepMIMO scenarios, i2\_28b and boston5g\_28, as shown in Fig.~\ref{simulation}. The i2\_28b scenario represents an indoor environment in which all users are non-line-of-sight relative to the BS, whereas boston5g\_28 captures a more complex outdoor urban environment in Boston. In both scenarios, the BS operates at 28~GHz and is equipped with a 64-element uniform linear array of isotropic antennas with half-wavelength spacing.

We evaluate performance in terms of normalized beamforming gain versus the number of probing beams. In the multiple-generation version of GenSSBF, the model produces a set of candidate beams (set to five in this experiment) conditioned on the coarse RSRP vector, and then selects the best one according to the workflow shown in Fig.~\ref{workflow}. By contrast, the single-generation version directly outputs one beam without a selection stage. We compare the proposed method with three baselines: (i) \textbf{Optimal}, the upper bound with perfect CSI; (ii) \textbf{Discriminative regression}, a multilayer perceptron (MLP) trained to predict beamforming weights; and (iii) \textbf{DFT beams}, codebook-based sweeping using a fixed DFT codebook.

As shown in Fig.~\ref{simulation}, GenSSBF is particularly advantageous in the low-overhead regime where only a few probing beams are available. In the i2\_28b scenario, the fully non-line-of-sight environment introduces substantial ambiguity, so that a short probing-beam RSRP vector may correspond to multiple plausible propagation states and beam directions. Under this challenging setting, both the single- and multiple-generation versions of GenSSBF remain close to the optimal benchmark even with as few as 10 probing beams, whereas discriminative regression degrades sharply because its single deterministic prediction tends to average conflicting beam directions and thus fails to capture the underlying multimodal structure. The benefit of generating multiple candidate beams is further evidenced by the significant gain of multiple-generation GenSSBF over its single-generation counterpart. This implies that by producing several plausible beams and selecting the best one, the generate-and-select workflow more effectively resolves ambiguity from coarse observations. Similar trends are observed in the outdoor boston5g\_28 scenario, where DFT beams suffer from limited codebook resolution and beam misalignment, while GenSSBF maintains high normalized beamforming gain and clearly outperforms the discriminative baseline. The beampattern comparisons in Fig.~\ref{simulation} further corroborate these results, showing that GenSSBF produces beams much closer to the optimal pattern, whereas discriminative regression yields more distorted and lower-gain beams and DFT beams remain restricted to fixed grid directions.

\vspace{-0.3cm}
\section{Conclusion and Outlook}

This article proposed GenSSBF, a generative site-specific beamforming framework that treats beamforming as a multimodal structured prediction problem. By transitioning from discriminative point estimates to generative distributional modeling, GenSSBF helps resolve the ambiguity inherent in low-overhead channel sensing. However, as GenSSBF and the broader SSBF paradigm remain in the early stages of research, several open challenges and opportunities need to be addressed before they can be fully deployed in next-generation networks, as elaborated below.

\textbf{Multi-User Multi-Cell Optimization.} Current SSBF research primarily focuses on single-user beamforming within a single cell. However, operational networks must serve multiple users simultaneously while managing inter-cell interference. Extending GenSSBF to multi-user multi-cell scenarios requires modeling the joint conditional distribution of beamformers across distributed BSs. By learning to sample from this high-dimensional joint distribution, a cooperative GenSSBF agent could implicitly perform interference management and maximize network-level spectral efficiency, effectively acting as a real-time neural surrogate for computationally intensive iterative optimization algorithms.

\textbf{Scalability and Generalization.} A significant challenge in site-specific learning is the dependence on fixed input resolutions and specific site geometries, which limits the transferability of trained models to new deployments. To address this, the integration of Fourier neural operators (FNOs) \cite{li2021fourier} presents a promising direction for resolution-invariant learning. Unlike conventional networks that are tied to a specific number of antennas or a specific grid discretization, FNOs learn the continuous operator mapping between site geometry and wave propagation in the frequency domain, enabling a single GenSSBF model to generalize across varying antenna array sizes and digital twin resolutions without the need for extensive retraining.

\textbf{Distributed Multi-Site Training.} Training a unique generative model from scratch for every BS in a massive cellular network is computationally prohibitive and inefficient. Future implementations should leverage distributed multi-site training paradigms, such as federated learning, to collaboratively train a foundation model of radio propagation using data aggregated from diverse environments. This approach allows the network to learn universal physical features of signal interaction, such as diffraction and reflection patterns, while preserving local data privacy, ultimately enabling rapid few-shot adaptation where a pre-trained global model is fine-tuned to a new site with minimal overhead.

\vspace{-0.3cm}
\balance
\bibliographystyle{IEEEtran}
\bibliography{mybib}

\vspace{-1cm}

\begin{IEEEbiographynophoto} {Zhaolin Wang} is a Research Assistant Professor at The University of Hong Kong. His research interests include electromagnetic information theory, integrated sensing and communications, and artificial intelligence in wireless communications. He is the recipient of the 2023 IEEE Daniel E. Noble Fellowship Award and the 2025 IEEE CTTC Andrea Goldsmith Young Scholars Award. 
\end{IEEEbiographynophoto}
\vspace{-1cm}

\begin{IEEEbiographynophoto} {Zihao Zhou} is currently pursuing the Ph.D. degree at The University of Hong Kong. His research interests include next-generation intelligent wireless networks and deep learning in communications.
\end{IEEEbiographynophoto}
\vspace{-1cm}

\begin{IEEEbiographynophoto} {Cheng-Jie Zhao} is currently pursuing the Ph.D. degree at The University of Hong Kong. His research interests include signal processing, artificial intelligence, and their applications in wireless communications.
\end{IEEEbiographynophoto}
\vspace{-1cm}

\begin{IEEEbiographynophoto} {Yuanwei Liu} is a Professor at The University of Hong Kong and a Visiting Professor at Queen Mary University of London. His research interests include generative AI/LLM for communications, low altitude networks large model, mobile edge generation, wireless digital twins, and integrated sensing and communications. 
\end{IEEEbiographynophoto}


\end{document}